\documentclass[prl,aps,nofootinbib,superscriptaddress,showpacs,floatfix,preprintnumbers,twocolumn]{revtex4}
\usepackage{hyperref,amssymb,amsmath,graphicx,xcolor,slashed}

\DeclareMathOperator\Tr{Tr}

\begin{document}

\title{Nucleon Helicity Generalized Parton Distribution at Physical Pion Mass from Lattice QCD}

\author{Huey-Wen Lin}
\email{hwlin@pa.msu.edu}
\affiliation{Department of Physics and Astronomy, Michigan State University, East Lansing, MI 48824}
\affiliation{Department of Computational Mathematics,
  Science and Engineering, Michigan State University, East Lansing, MI 48824}

\preprint{MSUHEP-21-024}


\begin{abstract}

The generalized parton distributions (GPDs) offer a window on three-dimensional imaging of the nucleon, providing understanding of how the fundamental properties of the nucleon, such as its mass and spin, arise from the underlying quark and gluon degrees of freedom.
In this work, we present the first lattice calculation of the nucleon isovector helicity GPD at physical pion mass, using an $a \approx 0.09$~fm lattice ensemble with 2+1+1 flavors of highly improved staggered quarks generated by MILC Collaboration.
We perform the GPD calculation in Breit frame using averaged nucleon boost momentum $P_z \approx 2.2$~GeV with nonzero momentum transfers in $[0.2,1.0]\text{ GeV}^2$.
Nonperturbative renormalization in RI/MOM scheme is used to obtain the quasi-distribution before matching to the lightcone GPDs.
The three-dimensional distribution $\tilde{H}(x,Q^2)$
is presented, along with the three-dimensional nucleon tomography and impact-parameter--dependent distribution for selected Bjorken $x$ at $\mu=3$~GeV in $\overline{\text{MS}}$ scheme.
\end{abstract}

\maketitle

\section{Introduction}

Three-dimensional images of the nucleon are critically important for understanding the origin of mass and orbital angular momentum as well as other properties of the nucleon.
Generalized parton distributions (GPDs) use a single set of three-dimensional functions~\cite{Mueller:1998fv, Ji:1996ek} to describe the spatial and momentum structures of the nucleon, including in its limits the parton distributions functions (PDFs) and the elastic form factors (FFs).
Experimentally, GPDs can be accessed in exclusive processes such as deeply virtual Compton scattering or meson electroproduction~\cite{Ji:1996nm}.
Worldwide experimental collaborations and facilities have been devoted to searching for these last unknowns of the nucleon;
these include HERMES at DESY, COMPASS at CERN, J-PARC in Japan, Halls A, B, and C at Jefferson Laboratory, and PHENIX and STAR at RHIC (Brookhaven National Laboratory) in the US.
The pursuit of GPDs and their imaging has led the hadronic-physics community worldwide to plan future experiments:
the Electron-Ion Collider~\cite{Accardi:2012qut},
Electron-Ion Collider in China (EicC)~\cite{Anderle:2021wcy,AbdulKhalek:2021gbh},
and the Large Hadron-Electron Collider (LHeC) in Europe~\cite{AbelleiraFernandez:2012cc,Agostini:2020fmq}.
However, obtaining high-quality images from experimental data remains decades away.
Meanwhile, large-scale numerical lattice-QCD calculation can fill in some of the gaps where experiments cannot reach, such as zero-skewness limit.
In this work, we focus on the longitudinally polarized GPD functions $\tilde H(x, \xi, t)$.

The two helicity (longitudinally polarized) GPD functions $\tilde H(x, \xi, t)$ and $\tilde E(x, \xi, t)$ are defined in terms of the matrix elements~\cite{ Ji:1996ek}
\begin{widetext}
\begin{align}
\tilde F_q(x,\xi,t)&=\int\frac{dz^-}{4\pi}e^{ixp^+ z^-}\left\langle p''\left|\bar\psi\left(-\frac{z}{2}\right)\gamma^+ \gamma^5 L\left(-\frac{z}{2},\frac{z}{2}\right)\psi\left(\frac{z}{2}\right)\right|p'\right\rangle_{z^+=0,\vec z_\perp=0}\nonumber\\
&=\frac{1}{2p^+}\left[\tilde H(x,\xi,t)\bar u(p'')\gamma^+ \gamma^5 u(p')+\tilde E(x,\xi,t)\bar u(p'')\frac{\gamma^5\Delta^+}{2m}u(p')\right],
\end{align}
\end{widetext}
where $\Delta^\mu=p''^\mu-p'^\mu$, $t=\Delta^2$, skewness $\xi=\frac{p''^+-p'^+}{p''^+ +p'^+}$, and the
gauge link lies along the lightcone $L(-z/2,z/2)$.
In the zero-skewness limit $\xi, \to 0$, $\tilde{H}$ reduces to the usual helicity parton distributions $\Delta q (x)$.

Large-momentum effective theory (LaMET), also known as the ``quasi-PDF method''~\cite{Ji:2013dva,Ji:2014gla,Ji:2017rah}, allows us to  connect quantities calculable on the lattice to those on the lightcone, and using lattice results to probe the full Bjorken-$x$ dependence of distributions for the first time. 
Since then, there have been many lattice works calculated on nucleon and meson PDFs based on the quasi-PDF approach~\cite{Lin:2013yra,Lin:2014zya,Chen:2016utp,Lin:2017ani,Alexandrou:2015rja,Alexandrou:2016jqi,Alexandrou:2017huk,Chen:2017mzz,Alexandrou:2018pbm,Chen:2018xof,Chen:2018fwa,Alexandrou:2018eet,Lin:2018qky,Fan:2018dxu,Liu:2018hxv,Wang:2019tgg,Lin:2019ocg,Chen:2019lcm,Lin:2020reh,Chai:2020nxw,Bhattacharya:2020cen,Lin:2020ssv,Zhang:2020dkn,Li:2020xml,Fan:2020nzz,Gao:2020ito,Lin:2020fsj,Zhang:2020rsx,Alexandrou:2020qtt,Alexandrou:2020zbe,Lin:2020rxa,Gao:2021hxl,Lin:2020rut,Chen:2018fwa,Sufian:2019bol,Izubuchi:2019lyk,Joo:2019bzr,Sufian:2020vzb,Shugert:2020tgq,Gao:2020ito}.
Alternative approaches to lightcone PDFs in lattice QCD are
``operator product expansion (OPE) without OPE''~\cite{Aglietti:1998ur,Martinelli:1998hz,Dawson:1997ic,Capitani:1998fe,Capitani:1999fm,Chambers:2017dov,QCDSF-UKQCD-CSSM:2020tbz,Horsley:2020ltc},
“auxiliary heavy/light quark”~\cite{Detmold:2005gg,Detmold:2018kwu,Detmold:2020lev,Braun:2007wv},
“hadronic tensor”~\cite{Liu:1993cv,Liu:1998um,Liu:1999ak,Liu:2016djw,Liu:2017lpe,Liu:2020okp},
``good lattice cross sections''~\cite{Ma:2017pxb,Bali:2017gfr,Bali:2018spj,Sufian:2019bol,Sufian:2020vzb} and the pseudo-PDF approach~\cite{Orginos:2017kos,Karpie:2017bzm,Karpie:2018zaz,Karpie:2019eiq,Joo:2019jct,Joo:2019bzr,Radyushkin:2018cvn,Zhang:2018ggy,Izubuchi:2018srq,Joo:2020spy,Bhat:2020ktg,Fan:2020cpa,Sufian:2020wcv,Karthik:2021qwz}.
There have been some initial $x$-dependent GPD studies on the lattice recently based on the  quasi-GPD~\cite{Ji:2015qla,Liu:2019urm} approach:
ETM Collaboration used LaMET method to calculate both unpolarized and polarized nucleon isovector GPDs with largest boost momentum 1.67~GeV at pion mass $M_\pi \approx 260$~MeV~\cite{Alexandrou:2020zbe} with one momentum transfer.
MSULat also reported the first lattice-QCD calculation of the unpolarized nucleon GPD with boost momentum around 2.0~GeV at the physical pion mass with multiple transfer momenta, allowing study of the three-dimensional structure and impact-parameter--space distribution~\cite{Lin:2020rxa}.
In this work, we report the result of the helicity GPDs at physical pion mass.

\section{Lattice Matrix Elements}

\begin{figure}[tb]
\includegraphics[width=0.45\textwidth]{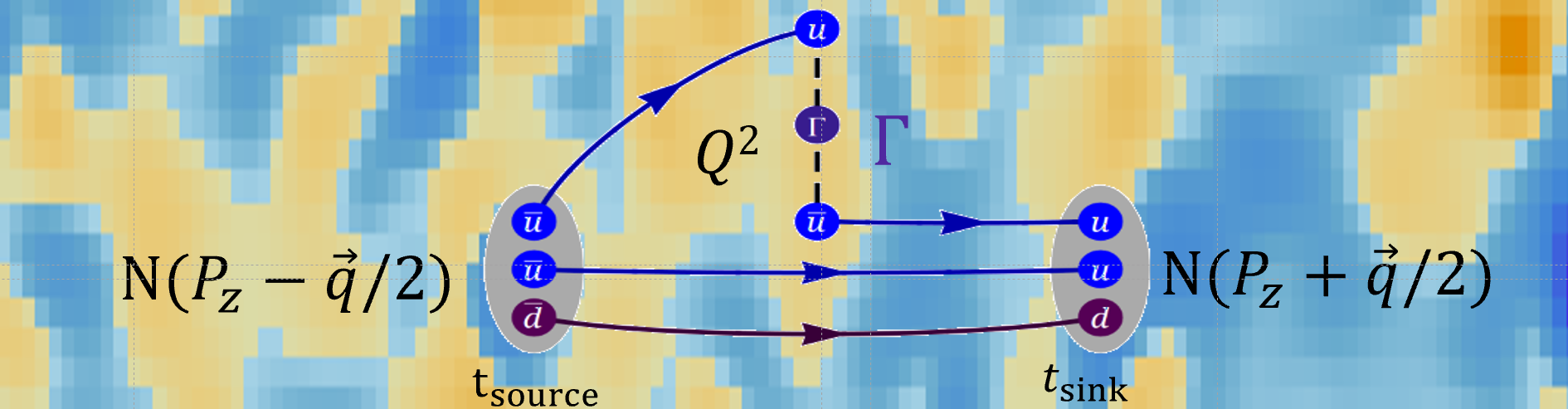}
\caption{Illustration of the Breit-frame lattice matrix-element calculation on top of the QCD vacuum.
\label{fig:GPD-diagram}}
\end{figure}

In this work, we focus on the nucleon isovector polarized GPDs and their quasi-GPD counterparts defined in terms of spacelike correlations calculated in Breit frame.
We use an ensemble at physical pion mass with $N_f=2+1+1$ (degenerate up/down, strange and charm) flavors of highly improved staggered dynamical quarks (HISQ)~\cite{Follana:2006rc} generated by MILC Collaboration~\cite{Bazavov:2012xda}.
The ensemble has lattice spacing $a\approx 0.09$~fm, and four-dimensional volume of $64^3\times 96$.
We use one step of hypercubic (HYP) smearing~\cite{Hasenfratz:2001hp} on the gauge links to suppress discretization effects.
We use clover valence fermion action with the clover parameters tuned to recover the lowest sea pion mass of the HISQ quarks.
The same ``mixed-action'' parameter choices have been used in other lattice calculations of nucleon charges, moments and form factors~\cite{Mondal:2020cmt,Jang:2019jkn,Gupta:2018lvp,Lin:2018obj,Gupta:2018qil,Rajan:2017lxk,Yoon:2016jzj,Bhattacharya:2016zcn,Bhattacharya:2015esa,Bhattacharya:2013ehc,Briceno:2012wt,Bhattacharya:2011qm,Lin:2020reh},
and there has been promising agreement between the calculated quantities and the experimental data when applicable.
We use Gaussian momentum smearing~\cite{Bali:2016lva} on the quark field to improve the overlap with the boosted-momentum ground-state nucleon.

We calculate matrix elements of the form $\left\langle N(P_f)\left|\bar\psi\left(-\frac{z}{2}\right)\Gamma L\left(-\frac{z}{2},\frac{z}{2}\right)\psi\left(\frac{z}{2}\right)\right|N(P_i)\right\rangle$ with projection operators $\Gamma = \frac{1+\gamma_t}{2}(1+i \gamma_5\gamma_{x,y,z})$.
An illustration of our Breit-frame setup can be found in Fig.~\ref{fig:GPD-diagram}.
To calculate the GPD matrix elements at nonzero momentum transfer, we first calculate the matrix element $\langle \chi_N (\vec{P}_f) | O^\mu | \chi_N (\vec{P}_i) \rangle$, where $\chi_N$ is the nucleon spin-1/2 interpolating field, $\epsilon^{abc} [q^{a\top}(x)C\gamma_5q^b(x)]q^c(x)$.
$O_\mu=\overline{\psi}\gamma_\mu W(z) \psi$ is the LaMET Wilson-line displacement operator with $\psi$ being either an up or down quark field, and $\vec{P}_{\{i,f\}}$ are the initial and final nucleon momenta.
We integrate out the spatial dependence and project the baryonic spin, using projection operators $\mathbb{P}_\rho=\frac{1+\gamma_t}{2}(1+i \gamma_5\gamma_{\rho})$ with $\rho\in\{x,y,z\}$, leaving a time-dependent three-point correlator, $C_\text{3pt}$.
\begin{widetext}
\begin{eqnarray}\label{eq:general-3pt}
\Gamma^{(3),\mathbb{P}_\rho}_{\mu,AB}(t_i,t,t_f,\vec{p}_i,\vec{P}_f) &=&
Z_O \sum_n \sum_{n^\prime} f_{n,n^\prime}(P_f,P_i,E_n^\prime,E_n,t,t_i,t_f)\nonumber \\
&\times& \sum_{s,s^\prime}
(\mathbb{P}_\rho)_{\alpha\beta} u_{n^\prime}(\vec{P}_f,s^\prime)_\beta
\langle N_{n^\prime}(\vec{P}_f,s^\prime)\left|O_\mu\right|N_n(\vec{P}_i,s)\rangle\overline{u}_n(\vec{P}_i,s)_\alpha,
\end{eqnarray}
\end{widetext}
where $f_{n,n^\prime}(P_f,P_i,E_n^\prime,E_n,t,t_i,t_f)$ contains kinematic factors involving the energies $E_n$ and overlap factors $A_n$ obtained in the two-point variational method, $n$ and $n^\prime$ are the indices of different energy states and $Z_O$ is the operator renormalization constant (which is determined nonperturbatively).
We use high-statistics measurements, 501,760 total over 1960 configurations, to help with statistical noise at high boost momenta, $P_z=|\frac{\vec{P}_i+\vec{P}_f}{2}| = |\frac{2\pi}{L}\{0,0,10\}|$ with $L=5.63$~fm.
We vary the spatial momentum transfer $\vec{q}=\vec{P}_f-\vec{P}_i=\frac{2\pi}{L}\{n_x,n_y,0\}$ with integer $n_{x,y}$ and $n^2=n_x^2+n_y^2 \in \{0,4,8,16,20\}$
with four-momentum transfer squared { $Q^2=-q_\mu q^\mu=\{{0, 0.19, 0.39, 0.77, 0.97}\} $}~GeV$^2$ using periodic boundary condition with all $|q|$ at fixed $Q^2$, rotationally averaged.
We set the quark momentum smearing parameter to $\{0,0,6\} \frac{2\pi}{L}$ for $n^2=\{0,4,8\}$, and $\{\pm 1 ,\pm 1 , 6 \}\frac{2\pi}{L}$ for $n^2=\{16,20\}$.
We simultaneous fit the 501,760-measurement three-point correlators with source-sink separation of $t_\text{sep} =[8,12]$ lattice units to extract the ground-state matrix elements. 
The details of how we extract nucleon ground-state matrix elements can be found in our previous works~\cite{Lin:2018obj,Lin:2018qky,Lin:2020rxa}.

\begin{figure*}[tb]
\includegraphics[width=0.42\textwidth]{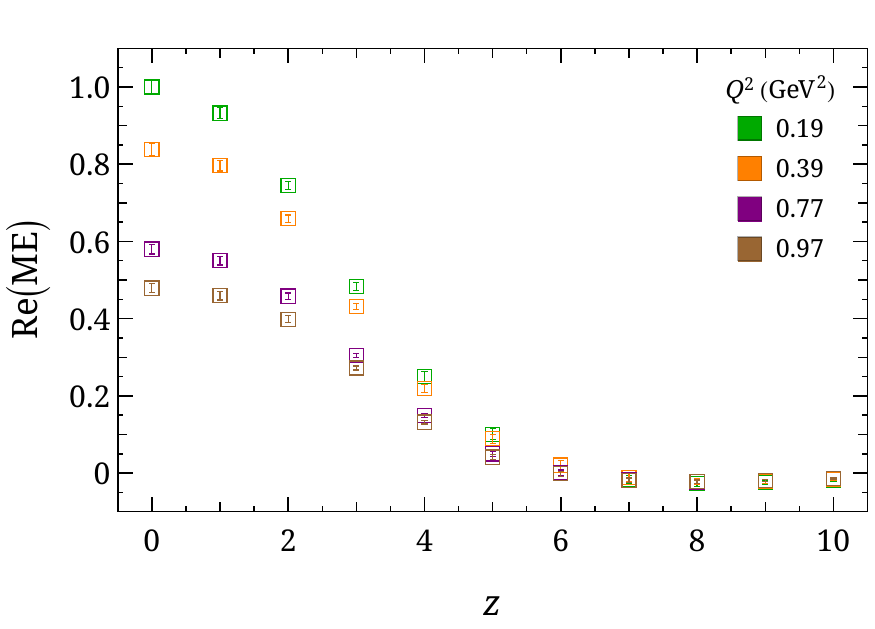}
\includegraphics[width=0.42\textwidth]{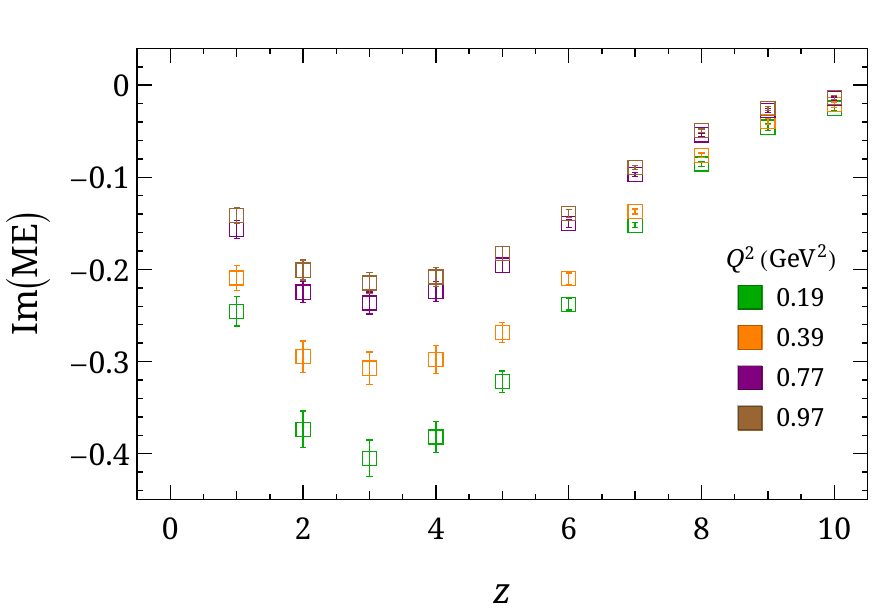}
\caption{
The Wilson-line length displacement $z$-dependence of matrix elements $h_{\tilde{H}}(P_z,Q^2,z)$ at selected momentum transfer $Q^2\in\{0.19, 0.39, 0.77, 0.97\}\text{ GeV}^2$.
\label{fig:H-MEs}}
\end{figure*}

The ground-state matrix elements are proportional to
\begin{widetext}
\begin{equation}
\label{eq:ffunc}
 \frac{  \Tr \left\{ \mathbb{P}_\rho \, [ -i \slashed{P}_{\!f} + m_N] \:
 [h_{\tilde{H}} (P_z,Q^2,z)\gamma_5\gamma_z +h_{\tilde{E}}( P_z,Q^2,z)\frac{\gamma_5 q_z}{2M_N}]
 \, [ -i \slashed{P}_{\!i} + m_N] \, \right\} }{4 \, E_N(\vec{P}_f) E_N(\vec{P}_i)}
\end{equation}
\end{widetext}
with $\slashed{p} = i E_N(\vec{p}\,) \gamma_4 + \vec{p} \cdot \vec{\gamma}$.
Because $h_{\tilde{E}}$ always couples with $q_z=(P_f-P_i)_z$, for fixed operator $\Gamma=\gamma_z\gamma_5$ and zero-skewness limit, the second term of the above equation will always be zero.
We do not need to solve a linear system of equations to get $\tilde{H}$  in coordinate space, $h_{\tilde{H}}(P_z,Q^2,z)$.
We then normalize $h_{\tilde{H}}(P_z,0,0)$ by $h_{\tilde{H}}(P_z,0,0) \times g_A^\text{exp}$ with $g_A^\text{exp} = 1.27$.
Selected normalized $Q^2$ values of $h_{\tilde{H}}(P_z,Q^2,z)$ are shown in Fig.~\ref{fig:H-MEs}. %
The real matrix elements decrease quickly to zero due to the large boost momentum used in this calculation.
Using large boost momentum helps to reduce the contributions from higher-twist effects at $O(\Lambda_\text{QCD}^2/P_z^2)$.

\section{Results on Lattice Helicity GPD}

Following recent work~\cite{Chen:2018xof,Lin:2018qky,Liu:2018hxv}, we nonperturbatively renormalize the matrix elements in RI/MOM scheme with $\mu_R=3.8$~GeV.
We then Fourier transform the renormalized matrix elements into quasi-GPDs through two approaches:
1)~We take the matrix elements $z \in [-12,12]$ and apply the simple but effective ``derivative'' method,
$\tilde{Q} = i\int_{-z_\text{max}}^{+z_\text{max}} \!\! dz\, e^{i x P_z z}  \tilde{h}'_R/x$, to obtain the quasi-GPDs.
2)~We adopt the extrapolation formulation suggested by Ref.~\cite{Ji:2020brr} by fitting $|z| \in \{10,15\}$ using the formula $c_1(-izP_z)^{-d_1}+c_2 e^{izP_z}(izP_z)^{-d_2}$, inspired by the Regge behavior, to extrapolate the matrix elements into the region beyond the lattice calculation and suppress Fourier-transformation artifacts.
Then, both quasi-GPDs are matched to the lightcone  GPDs by applying the matching condition~\cite{Liu:2018uuj,Chen:2018xof,Lin:2018qky}.

\begin{figure}[tb]
\centering
\includegraphics[width=0.4\textwidth]{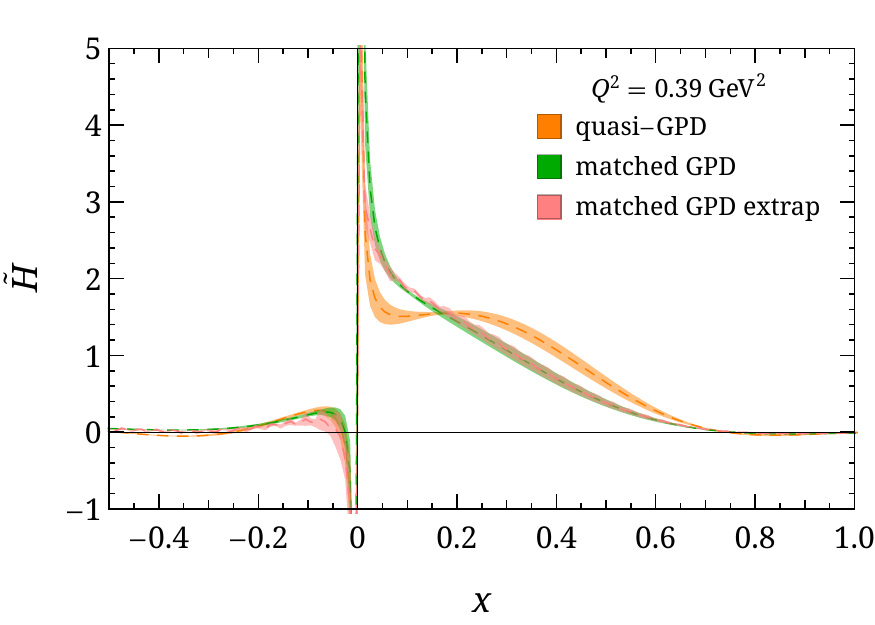}
\caption{
Nucleon isovector ${\tilde{H}}$ quasi-GPDs and lightcone GPDs at momentum transfer $Q^2=0.39\text{ GeV}^2$.
The orange and green bands are the quasi-GPD and lightcone GPDs from derivative method~\cite{Lin:2017ani}, while the pink band corresponds to the matched GPD using quasi-GPD from the extrapolation formulation suggested by Ref.~\cite{Ji:2020brr}.
We find both methods give reasonable agreement in the $x$-dependent behavior, except in the small-$x$ and negative-$x$ region, which is dominated by the large-$z$ matrix elements that rely on the extrapolation.
\label{fig:quasi-matched}}
\end{figure}

Figure~\ref{fig:quasi-matched} shows the quasi- and lightcone distribution of the ${\tilde{H}}$ GPD at momentum transfer $Q^2\approx 0.4\text{ GeV}^2$ using $P_z \approx 2.2$~GeV.
We find that quasi-GPD using both derivative and Regge-inspired extrapolation agree in the mid- to large-$x$ regions, but their difference grows as $x$ approaches zero in both the quark and antiquark distribution.
This is expected, since they differ mainly in the treatment of the large-$z$ matrix elements in the quasi-GPD Fourier transformation outside the region with available lattice data ($z>15$), which contributes more significantly to the small-$x$ distribution.
We also noticed some difference in the antiquark region ($x<0$):
this is also expected based on past work~\cite{Lin:2017ani,Chen:2018xof,Lin:2018qky,Liu:2018hxv} that even higher boost momenta are needed to improve the antiquark region.
We will take the difference between the two different ways of determining the quasi-GPD as a systematic reflected in the final uncertainties.
We also see that the matching lowers the positive mid-$x$ to large-$x$ quasi-distribution, as expected.
As one approaches the lightcone limit, the probability of a parton carrying a larger fraction of its parent nucleon's momentum should become smaller.
Note that the matching from quasi-GPD to GPD has residual  systematics at
$O\left(\frac{\Lambda_\text{QCD}^2}{(xP_z)^2}\right)$ and $O\left(\frac{\Lambda_\text{QCD}^2}{(1-x)^2P_z^2}\right)$ at very small $x$ and $x$ near 1.
We add $\frac{\Lambda_\text{QCD}^2}{P_z^2}$ into the systematic errors to estimate this effect, but future calculations including larger momenta will characterize this systematic better.

\begin{figure}[tb]
\includegraphics[width=0.4\textwidth]{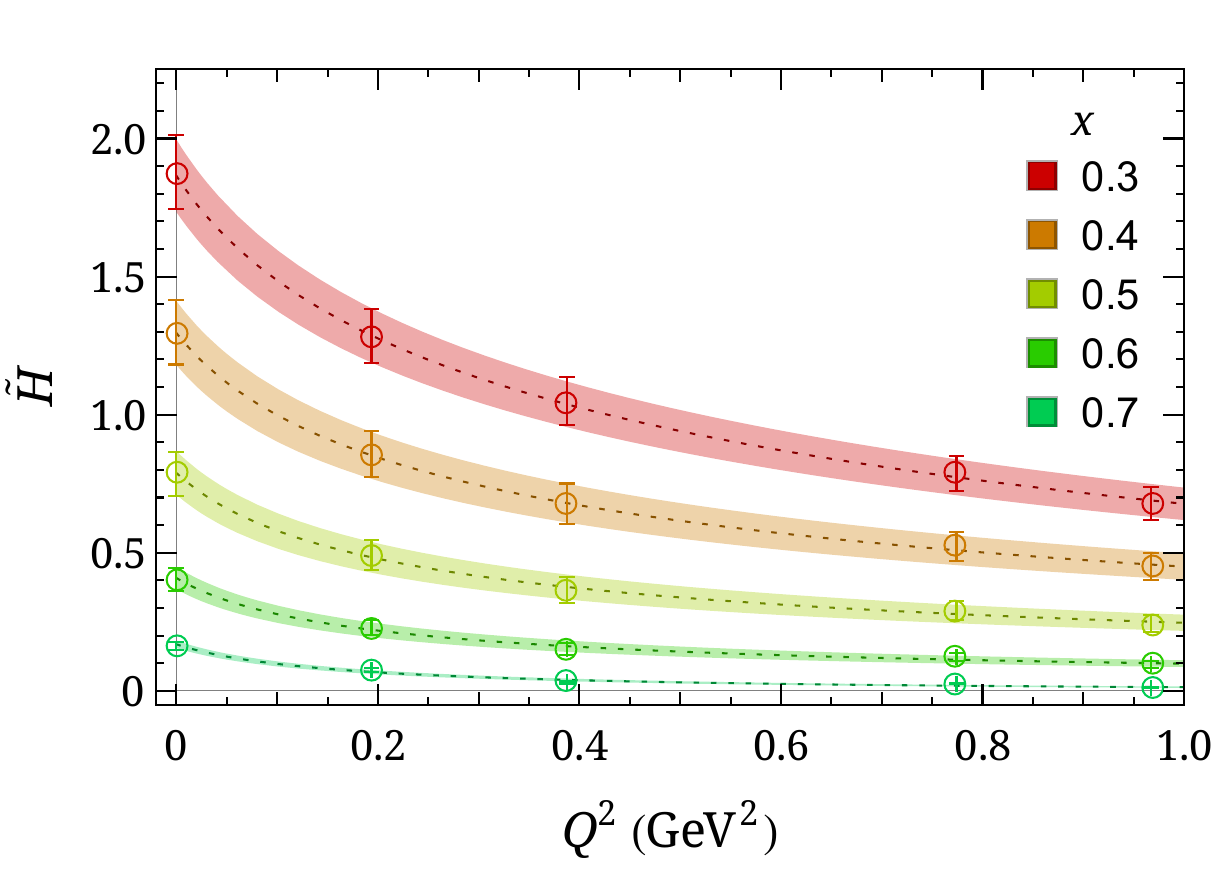}
\includegraphics[width=0.4\textwidth]{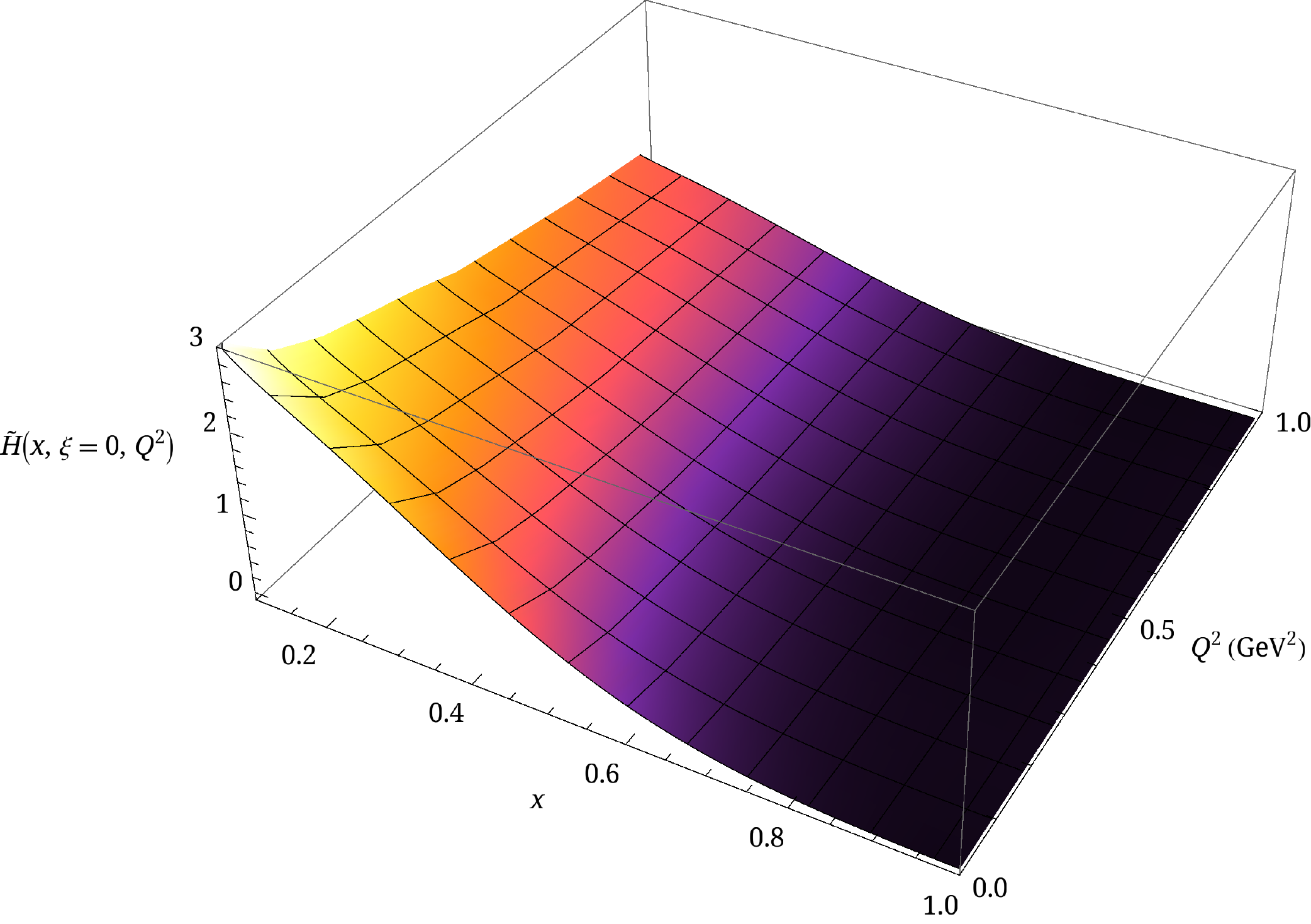}
\caption{
Nucleon isovector ${\tilde{H}}$ GPDs at $\xi=0$ with
$z$-expansion to $Q^2$ at selected $x$ values (top) and
 as functions of $x$ and momentum transfer $Q^2$ (bottom).
\label{fig:3DGPD}}
\end{figure}

For convenience, we will focus on showing the GPD results from the derivative method, and use the Regge-inspired extrapolation to estimate small-$x$ in reconstructing GPD moments from our $x$-dependent GPD functions.
For the rest of the work, we will mainly focus on the $x>0.05$ region.
We repeat a similar analysis for each available $Q^2$ in this calculation, then use $z$-expansion up to 3 parameters to interpret the $Q^2$ dependence of the lightcone GPD functions; selected $x$ results are shown in the top figure of Fig.~\ref{fig:3DGPD}.
Through this fit, we can construct the full three-dimensional shape of ${\tilde{H}}$ as functions of $x$ and $Q^2$, as shown in the bottom of Fig.~\ref{fig:3DGPD}.

The zero-skewness limit of the GPD is related to the Mellin moments by taking the $x$-moments~\cite{Ji:1998pc,Hagler:2009ni}:
\begin{align}
\label{eq:GFFs}
\int_{-1}^{+1}\!\!dx \, x^{n-1} \, \tilde{H}(x, \xi, Q^2)&=  \nonumber \\
\sum\limits_{i=0,\text{ even}}^{n-1} (-2\xi)^i \tilde{A}_{ni}(Q^2) &+ (-2\xi)^{n} \,  \tilde{C}_{n0}(Q^2)|_{n\text{ even}},
\end{align}
where the generalized form factors (GFFs) $\tilde{A}_{ni}(Q^2)$,
and $\tilde{C}_{ni}(Q^2)$ in the $\xi$-expansion on the right-hand side are real functions.
When $n=1$ ($n=2$), we get the axial form factors $G_A(Q^2) = \tilde{A}_{10}(Q^2)$ (GFFs $\tilde{A}_{20}(Q^2)$).
There have been a number of calculations of the Mellin moments of the GPDs on the lattice using local matrix elements through the operator product expansion (OPE), so we can make a comparison between LaMET results and moment methods.

\begin{figure}[tb]
\centering
\includegraphics[width=0.43\textwidth]{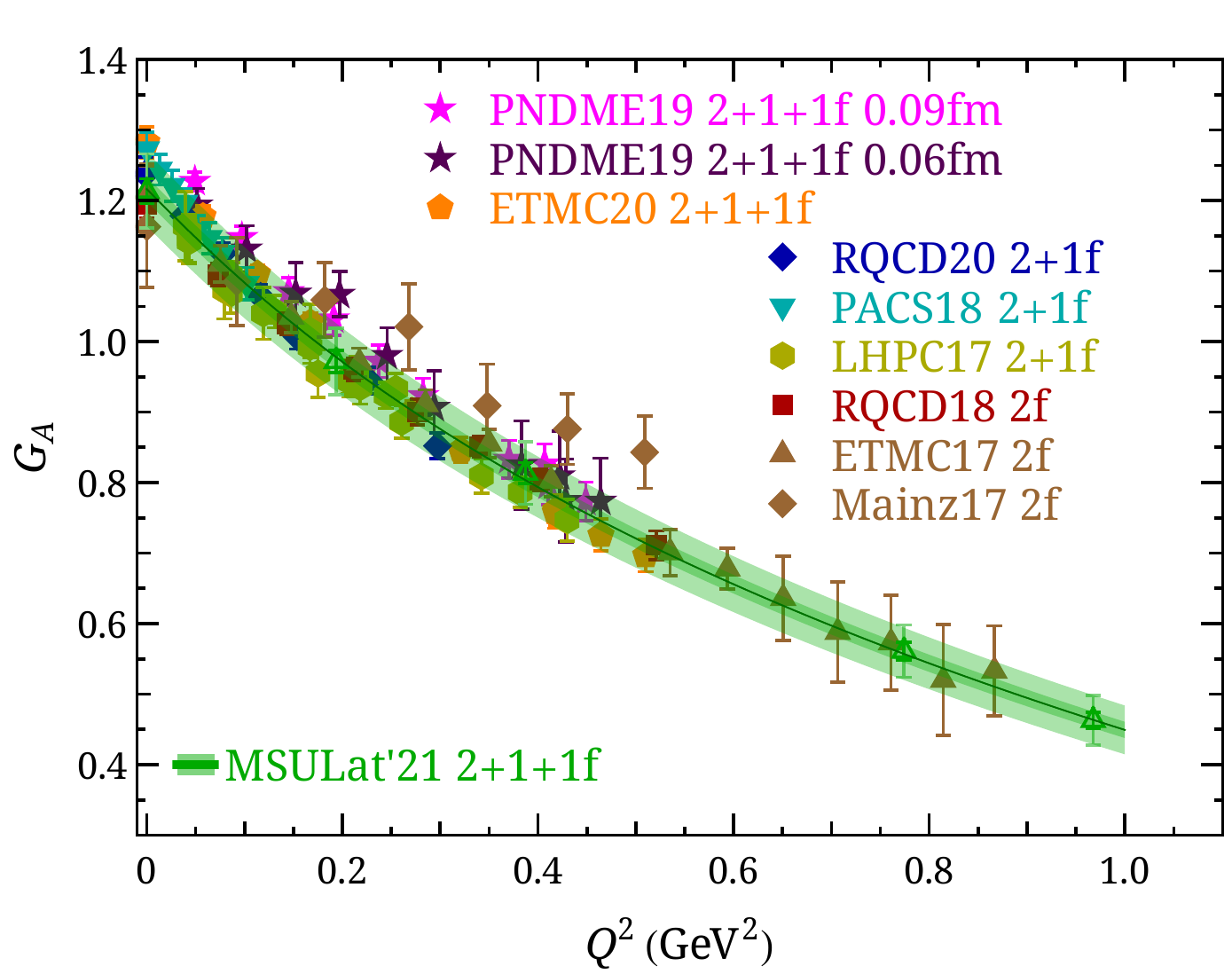}
\includegraphics[width=0.45\textwidth]{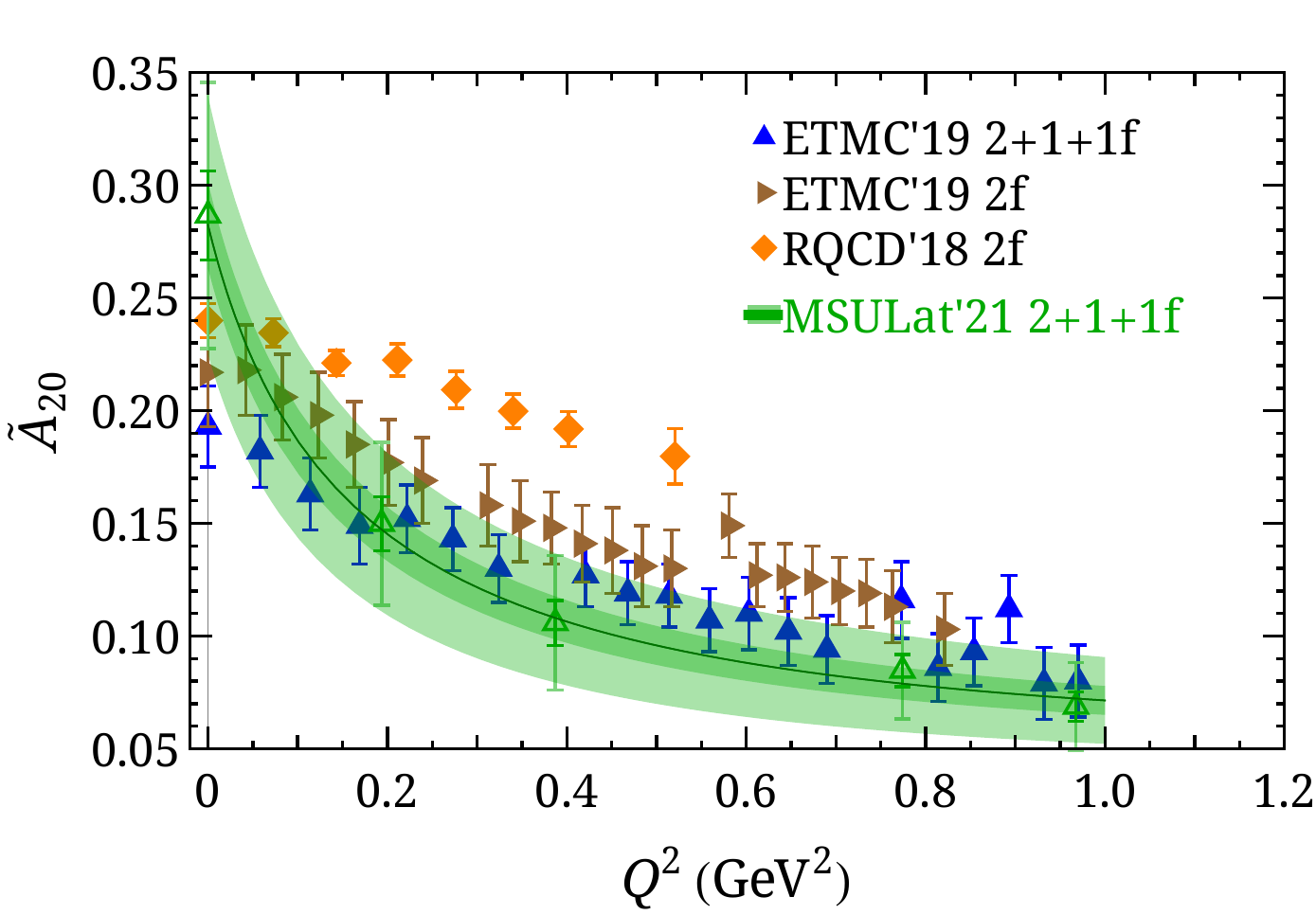}
\caption{
\label{fig:LatGFF}
(top) Nucleon isovector axial form factor as functions of $Q^2$ from other lattice calculations on ensembles near the physical pion mass~\cite{RQCD:2019jai,Alexandrou:2020okk,Capitani:2017qpc,Alexandrou:2017hac,Bali:2018qus,Hasan:2017wwt,Shintani:2018ozy,Rajan:2017lxk},
together with the form factor results obtained from this work (labeled as ``MSULat21 2+1+1'') by taking $n=1$ in Eq.~\ref{eq:GFFs}.
All the existing lattice calculations are done at a single lattice spacing, except for
PNDME19~\cite{Jang:2018djx}  with 2 lattice spacings of 0.06 and 0.09~fm, which we distinguish in the legend.
(bottom) The linearly polarized nucleon isovector GFF $\tilde{A}_{20}(Q^2)$ obtained from this work (labeled as ``MSULat21 2+1+1f'') by taking $n=2$ in Eq.~\ref{eq:GFFs}, compared with other lattice results calculated near physical pion mass as functions of transfer momentum $Q^2$:
2+1+1f ETMC19~\cite{Alexandrou:2019ali},
2f ETMC19~\cite{Alexandrou:2019ali} (only the larger-volume results are shown here),
2f RQCD19~\cite{Bali:2018zgl}.
}
\end{figure}

We show the results $\tilde{A}_{10}(Q^2)$ and $\tilde{A}_{20}(Q^2)$ obtained from this work (labeled as ``MSULat21 2+1+1'') as a function of momentum transfer $Q^2$ in Fig.~\ref{fig:LatGFF}, along with other lattice calculations near physical pion mass using traditional local-operator methods.
We use $z$-expansion to interpret the $Q^2$ dependence of our calculation, and the results are shown as the green band in Fig.~\ref{fig:LatGFF}.
The inner-band error indicates the statistical errors, while the outer one includes the systematic errors coming from several sources considered.
First, the difference between the quasi-GPD methods, discussed previously.
We also create pseudo-lattice data using the CT18NNLO PDF~\cite{Hou:2019efy} with the same lattice $z$ and $P_z$ parameters used in this calculation and take the upper limit of the reconstructed and original CT18 moments as an estimate of the systematics introduced by the analysis procedure (e.g. by Fourier truncation).
These steps should account for the lack of sensitivity to the negative- and small-$x$ regions in the current GPD extraction.
We vary the maximum Wilson-line length $z$ by 2 lattice units and take half the difference as an estimate of the systematic due to finite $z$ used in obtaining the quasi-GPD.
We also include an estimate of $O(1/P_z)$ systematics due to potential higher-twist effects by comparing our $Q^2=0$ PDFs to those in the previous works with 3 boost momenta~\cite{Chen:2018xof,Lin:2018qky}.
The finite-volume effects studied in our past work using the  HISQ ensembles~\cite{Lin:2019ocg} and an independent analytic study using ChPT~\cite{Liu:2020krc} were found to be smaller than typical statistical errors;
therefore, these are not considered in this work, given our $M_\pi L$ is close to 4.
The final errors are summed in quadrature to create the final error bands shown in Fig.~\ref{fig:LatGFF}.

We found our axial form factors, obtained from the ${\tilde{H}}$ GPD using LaMET method has slightly higher central values compared with other single--lattice-spacing lattice calculations, ranging from 2-flavor to 2+1+1-flavor ones using the axial-current approach with various choices of fermion and gauge action.
However, overall, it is consistent with past lattice calculations, whose results are shown with statistical errors only.
In the bottom plot of Fig.~\ref{fig:LatGFF}, we compare our moment results for $\tilde{A}_{20}(Q^2)$ with those obtained from simulations at the physical point by ETMC~\cite{Alexandrou:2019ali} and RQCD~\cite{Bali:2018zgl} using the OPE approach.
We note that even with the same OPE approach on $\tilde{A}_{20}(Q^2)$, lattice results from a single lattice spacing with different actions do not agree;
there have been indications that the systematic uncertainties are more complicated for these GFFs than those obtained for local currents, such as axial form factors.
The OPE operators are only expected to give the same results after taking the continuum $a \rightarrow 0$ limit, and current results only show a single lattice spacing.
Nevertheless, we find our results to be consistent with ETMC's 2+1+1-flavor results; this is perhaps a coincidence, since both our results and ETMC's are done using a single ensemble.
This first lattice calculation of the full three-dimensional $x$ and $Q^2$ dependence of the ${\tilde{H}}$ GPD functions using LaMET approach has a nice agreement with the previous moment approaches to the generalized form factors using using the OPE.

\begin{figure*}[tb]
\includegraphics[width=0.3\textwidth]{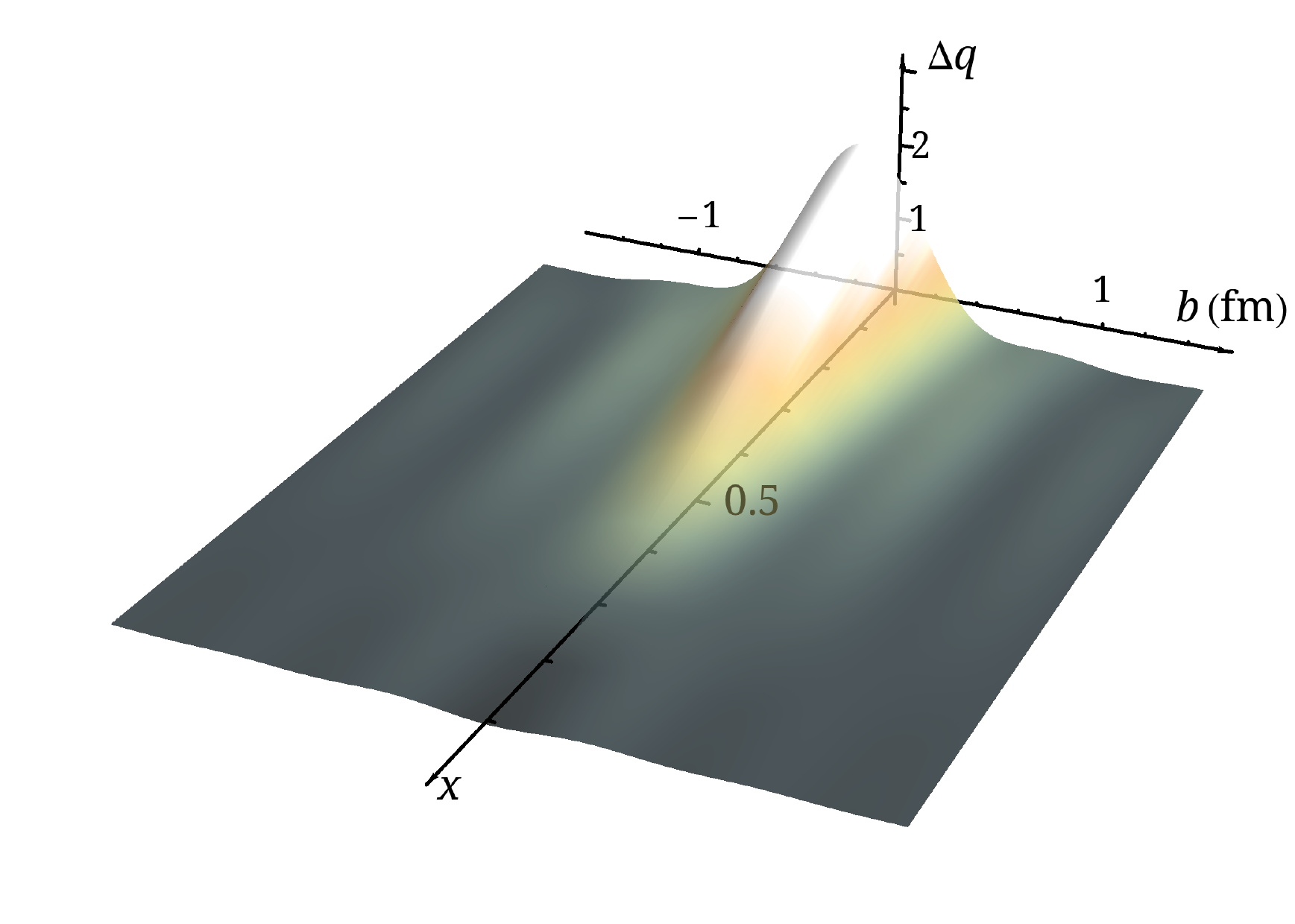}
\includegraphics[width=0.6\textwidth]{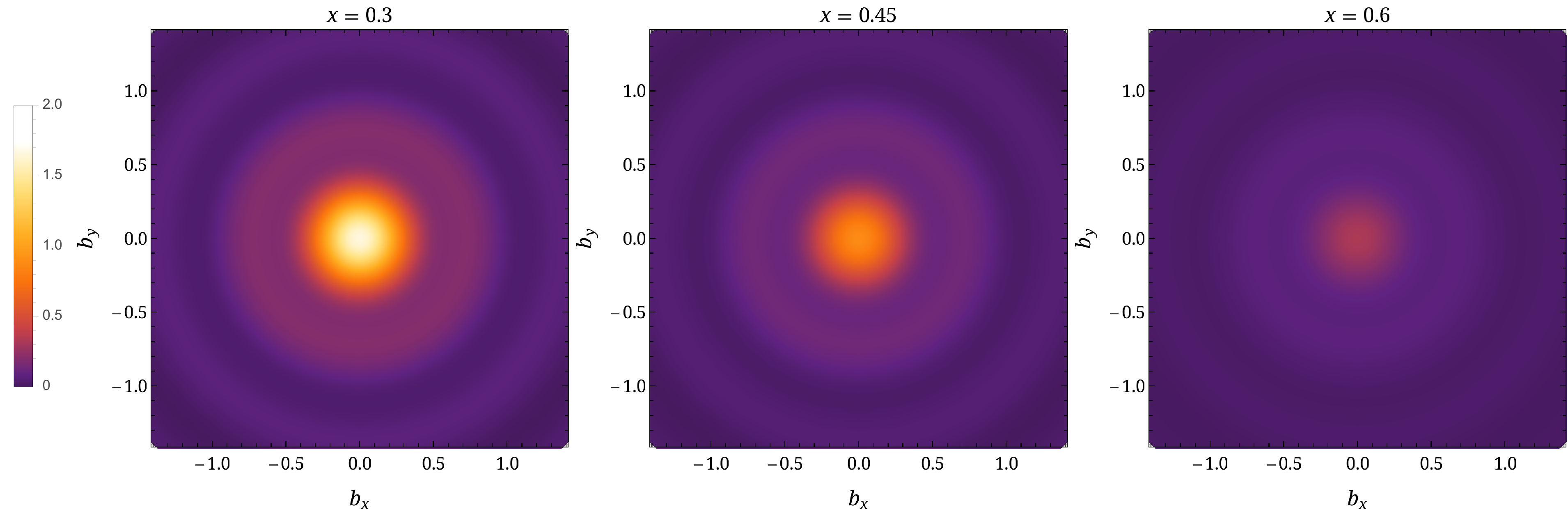}
\caption{
The impact-parameter--dependent polarized quark distribution $\mathsf{\Delta q}(x,b)$~\cite{Burkardt:2002hr} as a three-dimensional function of $x$ and $b$ (left) and two-dimensional distributions for $x=0.3$, 0.45 and 0.6.
These are obtained from our lattice $\tilde{H}$ at physical pion mass.
\label{fig:impact-distribution}}
\end{figure*}

The impact-parameter--dependent polarized quark distributions $\mathsf{\Delta q}(x,b)$~\cite{Burkardt:2002hr} can be obtained from ${\tilde{H}}$ using
\begin{equation}\label{eq:impact-dist}
\mathsf{\Delta q}(x,b) = \int \frac{ d \mathbf{q}}{(2\pi)^2} \tilde{H}(x,\xi=0,t=-\mathbf{q}^2) e^{i\mathbf{q}\,\cdot \, \mathbf{b} },
\end{equation}
where $b$ is the transverse distance from the center of momentum.
$\mathsf{\Delta q}(x,b)$ has only been obtained from QCD models so far;
this is the first lattice-QCD determination, shown in Fig.~\ref{fig:impact-distribution}.
We show the three-dimensional distribution as a function of $x$ and $b$, and two-dimensional distributions at $x=0.3$, 0.45 and 0.6 in
Fig.~\ref{fig:impact-distribution}.
It describes the probability density for a polarized parton with momentum fraction $x$ at distance $b$ in the transverse plane.
Compared with the $\mathsf{q}(x,b)$ obtained from the unpolarized GPD $H$ function from earlier work done on the same ensemble, the center of the two-dimensional distribution $\mathsf{q}(x,b=0) \leq \mathsf{\Delta q}(x,b=0)$ over most of the range of $x$.
The probability densities decrease quickly as $x$ and $b$ increase.

\section{Conclusion and Outlook}

In this work, we compute the zero-skewness isovector nucleon polarized  $\tilde{H}$ GPDs at physical pion mass using boost momentum $2.2$~GeV with nonzero momentum transfers in $[0.2,1.0]\text{ GeV}^2$.
We use two different methods to construct the quasi-GPD before matching to its lightcone counterpart.
We noted some differences using these two methods, and they have been taken account into systematics in the final results.
We also included the systematics from the choices of max Wilson-line displacement used in the LaMET operators and estimate the single boosted momentum dependence from previous helicity PDF studies.
When taking the integral of our $\tilde{H}$ in Eq.~\ref{eq:GFFs}, we are able to compare with previous calculations of the axial form factor $G_A$ and $\tilde{A}_{20}$ GFF with single ensembles near physical pion mass, since this is the first lattice $\tilde{H}$ calculation at physical pion mass.
We found consistent results with most of the past work within two sigma from various actions and other lattice parameters.
This seems to verify that LaMET GPD calculation gives reasonable results for pursuing precision calculations in the future.
We are able to map out the three-dimensional GPD $\tilde{H}$ structures using lattice QCD data for the first time.
Future work will investigate ensembles with smaller lattice spacing to reach even higher boost momentum so that we can push toward reliable determination of the smaller-$x$ and antiquark regions, adopt the hybrid renormalization scheme to improve the long-distance nonperturbative behavior~\cite{Ji:2020brr},
and expand the current study into the $\xi \neq 0$ GPDs.

\section{Acknowledgments}

We thank the MILC Collaboration for sharing the lattices used to perform this study. The LQCD calculations were performed using the Chroma software
suite~\cite{Edwards:2004sx,Osborn:2010mb,Babich:2010qb}.
This research used resources of the National Energy Research Scientific Computing Center, a DOE Office of Science User Facility supported by the Office of Science of the U.S. Department of Energy under Contract No. DE-AC02-05CH11231 through ERCAP;
facilities of the USQCD Collaboration, which are funded by the Office of Science of the U.S. Department of Energy,
and supported in part by Michigan State University through computational resources provided by the Institute for Cyber-Enabled Research (iCER).
The work of HL is partially supported by the US National Science Foundation under grant PHY 1653405 ``CAREER: Constraining Parton Distribution Functions for New-Physics Searches''
and by the  Research  Corporation  for  Science  Advancement through the Cottrell Scholar Award ``Unveiling the  Three-Dimensional Structure of Nucleons''.

%

\end{document}